\documentstyle{mn}

\input{epsf}

\begin{document}
\title[
Faint galaxy counts as a 
function of morphological type in a hierarchical 
merger model
]
{
Faint galaxy counts as a 
function of morphological type in a hierarchical 
merger model
}

\author[C.M. Baugh {\it et al.}]
{C.M. Baugh, S. Cole \& C.S. Frenk 
 \\
University of Durham, 
Department of Physics, Science Laboratories, South Road, Durham DH1 3LE
}
\maketitle

\def\mpc {h^{-1} {\rm{Mpc}}}
\def\and  {\it {et al.} \rm}
\def\rmd {\rm d}

\begin{abstract}

The unprecedented resolution of the refurbished Wide Field and Planetary
Camera 2 (WFPC2) on the Hubble Space Telescope (HST) has led to major
advances in our understanding of galaxy formation. The high image quality
in the Medium Deep Survey and Hubble Deep Field has made it possible, for
the first time, to classify faint distant galaxies according to
morphological type.  These observations have revealed a large population of
galaxies classed as irregulars or which show signs of recent merger
activity. Their abundance rises steeply with apparent magnitude, providing
a likely explanation for the large number of blue galaxies seen at faint
magnitudes. We demonstrate that such a population arises naturally in a
model in which structure forms hierarchically and which is dynamically
dominated by cold dark matter. The number counts of irregular, spiral and
elliptical galaxies as a function of magnitude seen in the HST data are well
reproduced in this model.We present detailed predictions for the outcome of
spectroscopic follow-up observations of the HST surveys. By measuring the
redshift distributions of faint galaxies of different morphological types,
these programmes will provide a test of the hierarchical galaxy formation
paradigm and might distinguish between models with different cosmological
parameters.

\end{abstract}



\section{Introduction}

Counting galaxies as a function of flux to very faint limits is one of the
main tools for tracing the evolutionary history of the galaxy
population. In combination with spectroscopic redshift measurements for
brighter subsamples, this diagnostic has uncovered significant evolution in
the galaxy population at moderate lookback times, corresponding to redshift
$z\simeq 0.5$ 
(Lilly \and 1995, Ellis \and 1996).
This raises the possibility that
the onset of the process of galaxy formation itself may soon become
accessible to observations.

The traditional approach to interpreting number counts is based on a
retrospective calculation, in which the locally observed galaxy luminosity
function and morphological mix are taken as the starting point. {\it Ad
hoc} assumptions are then made regarding the time evolution of the
luminosity and number density of galaxies and the epoch of galaxy
formation. This approach is limited because it bypasses an explanation of
the physical processes that drive galaxy evolution and lacks a clear link
to current ideas on the formation of galaxies by the gravitational growth
of primordial fluctuations.

A more satisfactory way to interpret faint galaxy data is to use the
new methodology of semianalytic modelling of galaxy formation and evolution
developed over the past five years 
(White \& Frenk 1991, Lacey \& Silk 1991, Cole 1991, 
Lacey \and 1993, Kauffmann \and 1993, Cole \and 1994).
In this approach, physically motivated models are
constructed {\it ab initio}, starting from the power spectrum of primordial
density fluctuations predicted by specific theories of structure
formation. The current level of understanding of the dynamics of cooling
gas, star formation, feedback of energy into prestellar
gas and galaxy mergers is encoded into a few simple rules, often
expressed in the form of scaling laws.


We describe the semianalytic model used in this paper in Section 2; for a
detailed discussion of the technique, see Cole \and (1994), and for an update
of results see Frenk, Baugh and Cole (1996).  The incorporation of
morphological types into the model is described in full in Baugh, Cole,
Frenk (1996).  Our predicted counts are compared with HST observations in
Section 3, along with predictions for the results of spectroscopic
follow-up studies.

\section{Semianalytic scheme for galaxy formation}

\begin{figure*}

{\epsfxsize=14.truecm \epsfysize=9.truecm 
\epsfbox[-30 430 530 700]{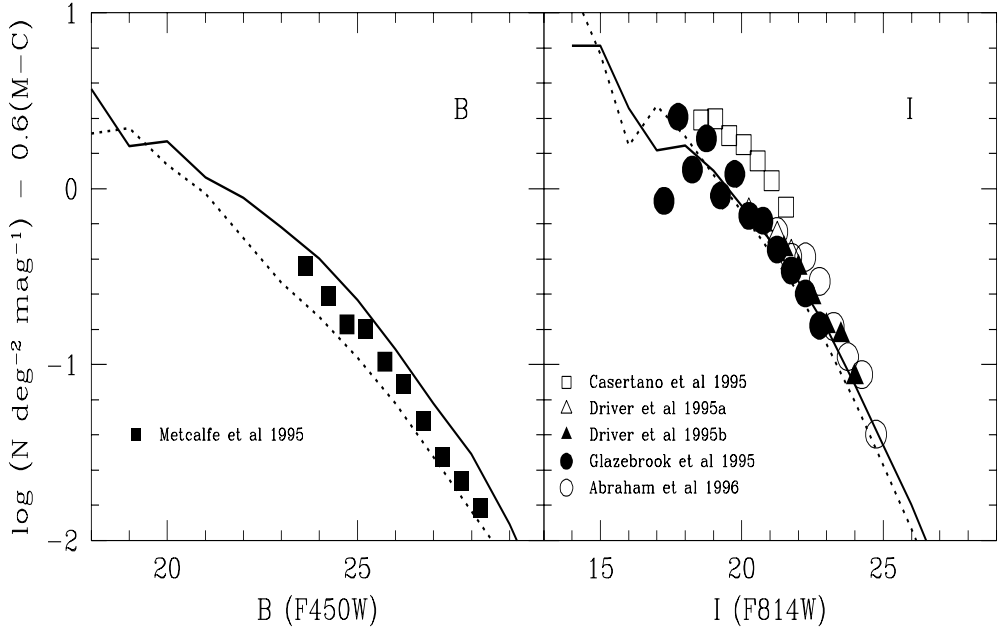}}
\caption[]
{
Number counts of galaxies of all types. The symbols show HST data 
 in the B-band  and I-band. 
The theoretical predictions are shown by the lines for two choices 
of IMF; the solid line shows a model with a Miller-Scalo IMF and the 
dotted line shows a model with a Scalo IMF.
The counts
have been divided by a power-law with the Euclidean slope of 0.6; the
constant $C=16$ for the B counts and $C=14$ for the I counts.  The
pre-refurbishment Medium Deep Survey (MDS) counts 
(open squares) are higher than the WFPC2
counts by up to $50 \%$.
}
\label{fig:hstall}
\end{figure*}

\begin{figure*}
{\epsfxsize=14.truecm \epsfysize=14.truecm 
\epsfbox[-75 160 580 700]{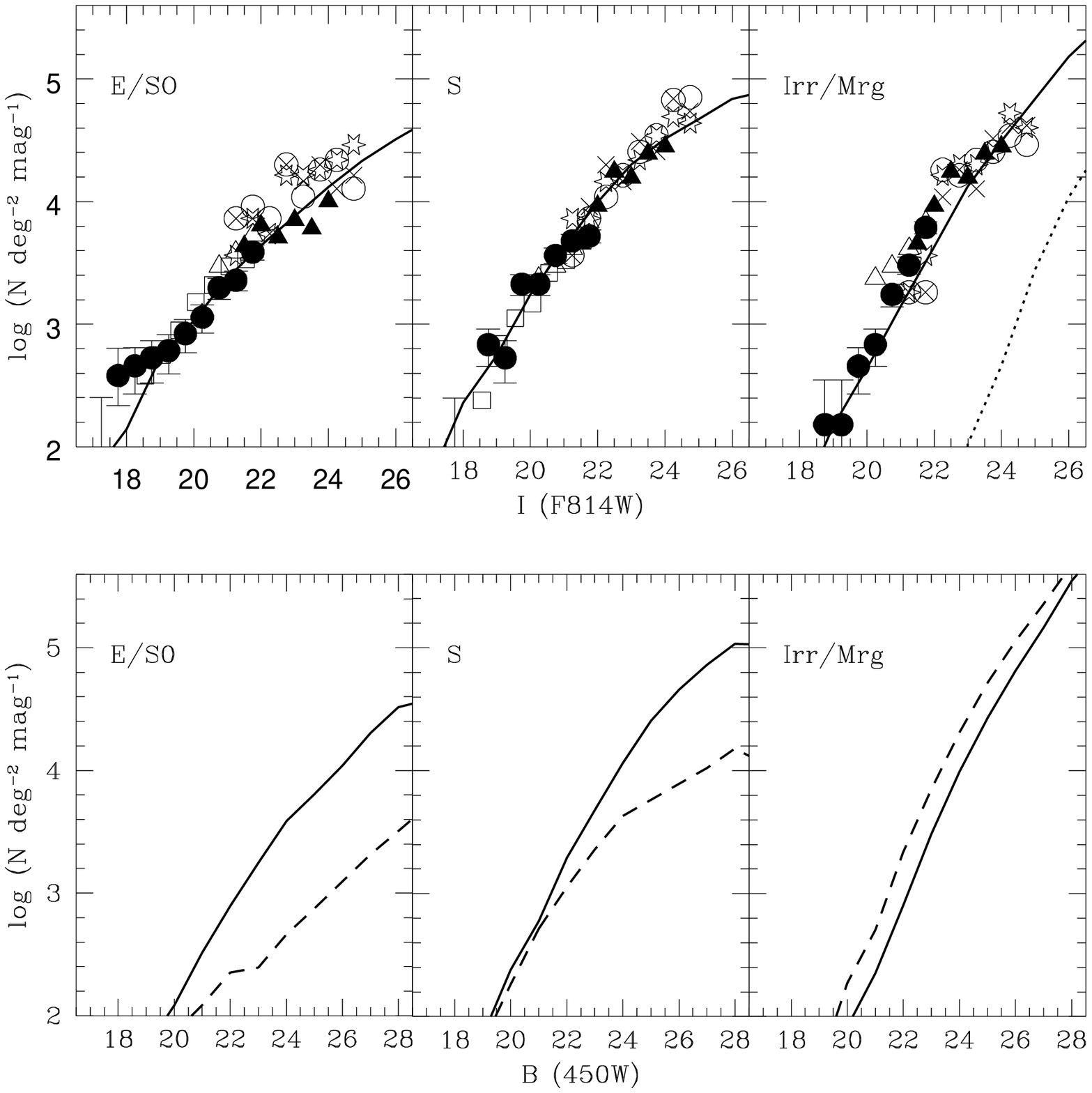}}
\caption[]
{
Number counts of galaxies of different morphological class. The upper
panels show I-band counts and the lower panels B-band counts. The symbols
represent I-band data from the same sources as in Figure~1, with the
addition of crosses and stars to represent the counts classified by eye
that supplement the automated counts in 
Abraham \and (1996a). The lines show our
theoretical predictions for a standard cold dark matter universe. The
different morphological classes are defined in terms of the bulge-to-disk
I-band luminosity ratio, as described in the text. The dotted line in the
Irr/Mrg panel shows the contribution to the counts from objects that have
undergone a major merger some time in the $1$ Gyr prior to
``observation". In the lower panels, the solid lines give the counts using
the morphological assignment derived from the I-band bulge-to-disk
ratio. The dashed line shows the counts when the galaxies are classified,
instead, by their bulge-to-disk B-band luminosity ratio (using the same
definitions of type employed in the I-band). 
}
\label{fig:hsttypes}
\end{figure*}

In the Cole {\it et al}  scheme, the hierarchical collapse and merging of an
ensemble of dark matter halos is followed using Monte-Carlo techniques
(Bond \and 1991; Cole \& Kaiser 1988).  Gas associated with a halo
virialises soon after halo collapse and then cools radiatively. Star
formation procedes at a rate proportional to the mass of cold gas.  The
amount of gas that cools is, in turn, regulated by feedback associated with
the effects of supernovae explosions and stellar winds.  When a merger of
dark matter halos occurs, the hot gas of the progenitors is stripped away
and assigned to the new halo. The cold gas and stars of the progenitors
typically merge on a longer timescale than the halos. This timescale and
other scaling parameters are calibrated by the results of numerical
simulations (Navarro, Frenk \& White 1993, 1995).  The luminosities of the
resulting galaxies are then calculated from their star formation histories
using stellar population synthesis models (Bruzual \& Charlot 1993; 1996 in
preparation; Charlot, Worthey \& Bressan 1996).  A surprisingly small
number of free parameters, five in all, is required to completely specify a
model of galaxy formation within a given cosmology.  Our general strategy
is to fix the values of the free parameters by attempting to match
properties of the {\it local} galaxy population as closely as
possible. This results in fully specified models that can then be used to
predict the properties of galaxies at high redshift. These models have been
successful in recovering the general form of the galaxy luminosity
function, the colours of galaxies and the counts and redshift distributions
of faint galaxies in the B and K bands (Cole \and 1994, Heyl \and 1995,
Frenk \and 1996).  However, a number of unresolved issues remain, most
notably the inability of these models to simultaneously reproduce the
observed local galaxy luminosity density and the zero-point of the
Tully-Fisher relation (White \& Frenk 1991; Kauffmann \and 1993; Cole \and
1994).

In this paper, we use an extension of the scheme of Cole {\it et al} in
which the light of each galaxy is separated into a bulge and a disk
component (Baugh, Cole \& Frenk 1996). 
The normal mode of star formation occurs quiescently
in a disk whilst the bulge component is assembled in galaxy
mergers. Residual star formation in the bulge may occur if a sufficiently
violent merger event triggers a burst of star formation. The 
bulge-to-disk ratio is thus a 
continually changing quantity and so the morphological type of a galaxy may
change depending upon whether a merger has just occured or whether there
has been a period of quiescent star formation.  This schematic
prescription is consistent with existing numerical simulations of galaxy
mergers 
(Barnes \& Hernquist 1992, Mihos \& Hernquist 1994a,b)
and with observations
of luminous IRAS galaxies (Clements \and 1996).

The ratio of the bulge-to-disk luminosity in the I-band, $(B/D)_I$, is used
to assign a broad morphological type to the model galaxies: ellipticals
have $(B/D)_I>0.65$, SOs $0.40<(B/D)_I<0.65$, spirals $0.10<(B/D)_I<0.40$,
and irregulars or late-type spirals $(B/D)_I<0.10$.  A galaxy that has
experienced a violent merger within the $1 {\rm Gyr}$ prior to
``observation" is classed as a ``merger" with a disturbed morphology. These
values of $(B/D)$ and the threshold mass (in the form of cold gas and
stars) that has to be accreted in a merger for it to be classed as violent
are set by requiring that the local morphological mix of galaxies in the
B-band be reproduced, within the uncertainties introduced by the
subjectiveness of morphological classification.  (Note that the 
observer-frame I-band
corresponds to the rest-frame B-band at a redshift of $z \sim 0.8$.)  As
discussed by Baugh \and (1996), this model produces a morphology-density
relation and a small scatter in the colour of cluster ellipticals similar
to those observed (Dressler 1980, Bower \and 1992), and gives rise to
``Butcher-Oemler" evolution in the morphological mix of cluster galaxies
(Butcher \& Oemler 1984).

In this paper, we use the fiducial parameters of the standard $\Omega=1$ 
cold dark matter model of Cole {\it et al}.
A full discussion of the effect of varying these
parameters or the underlying cosmology will be presented in a later paper
(Baugh, Cole \& Frenk in preparation). We set a dark matter halo 
mass resolution of $2 \times
10^{9} h^{-1} M_{\odot}$ in our Monte-Carlo scheme. Galaxy magnitudes are
computed using HST filters combined with the response of the optical
system. From the output of our model we construct a mock Hubble Deep Field
(HDF) catalogue, with magnitudes for each galaxy in the U(F300W),
B(F450W), V(F606W) and I(F814W) bands.

\section{Model Predictions}

\begin{figure}
{\epsfxsize=8.5truecm \epsfysize=12.truecm 
\epsfbox[25 160 580 750]{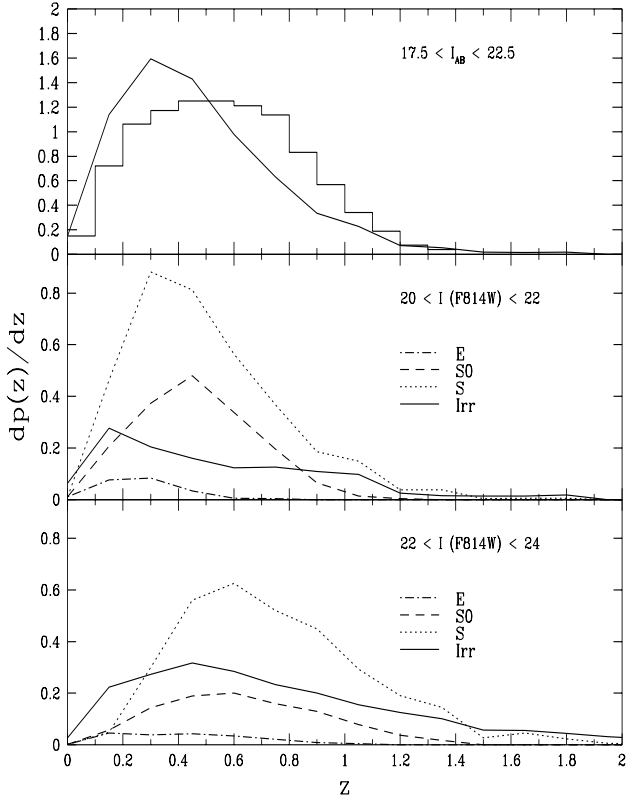}}
\caption[]
{
The redshift distribution of galaxy samples drawn from our mock HDF
catalogue. The top panel shows galaxies selected with the same apparent
magnitude limits as the Canada-France Redshift Survey data, which are
represented by the histogram from Lilly \and (1995). The central panel
gives our theoretical predictions for galaxies of different morphological
types and the same flux range as the sample selected for spectroscopic
follow-up by Driver {\it et al} (1995b). The area under each curve gives
the relative fraction of each galaxy type in the sample. The bottom panel
shows predictions for a fainter magnitude slice that could in principle be
observed with the Keck Telescope.
}
\label{fig:dndz}
\end{figure}


Figure \ref{fig:hstall} shows our predictions for the total counts in the
B- and I-bands, compared to a compilation of HST results.  
We have divided the counts by a power-law of Euclidian
slope, 0.6, in order to expand the ordinate. 
Our model predictions for the B-band counts depend somewhat on the assumed
IMF. 
Our predicted counts for two plausible choices (the Scalo (1986) and the
Miller-Scalo (Miller \& Scalo 1979) IMFs) bracket the HST data. 
In what follows we shall adopt the
Miller-Scalo IMF, but we note that the difference between the two sets
of predictions are only slightly larger than corrections ($\sim 0.5$ mag)
required to zero-point the data. 
The I-band predictions, on the other hand,
are very insensitive to the choice of IMF. 
The counts obtained from the
pre-refurbishment HST (open squares) are clearly inconsistent with the
WFPC2 data. 
Excluding these, our theoretical predictions are in excellent
agreement with the data. The split of the counts into different
morphological types is shown in Figure \ref{fig:hsttypes}. The HST galaxies
have been typed by eye by various groups (as indicated in the Figure
caption) and by an automated method that measures the asymmetry and central
concentration of the image (Abraham \and 1996a). In the top panels we compare
these HST I-band data with our model predictions which extend 1.5
magnitudes fainter than the data. At the faintest observed fluxes, $I\simeq
25$, the spiral and irregular/merger classes contain comparable numbers,
about a factor of 2 larger than the elliptical/SO class. Our model predicts
that the counts of E/SO and S galaxies begin to turn over beyond $I\simeq
25$, but the counts of irregular/merger galaxies continue to rise steeply
and dominates the I-band numbers faintwards of this limit. The main
contribution to this class comes from galaxies with small bulges. Our model
reproduces the observed trends remarkably well. By contrast, a
retrospective model based upon the locally observed fraction of irregulars,
a flat faint-end luminosity function and passive stellar evolution predicts
far fewer irregular/merger galaxies than observed at these faint limits
(Glazebrook \and 1995).

The lower panels of Figure \ref{fig:hsttypes} give our model predictions for the
counts in the B-band. The solid lines show the result of measuring
B-magnitudes for galaxies classified as above, {\it ie} on the basis of
their I-band bulge-to-disk ratio. Attempting to assign types using the
observer B-band bulge-to-disk ratio leads to less reliable results
(dashed lines in the bottom panel of Figure \ref{fig:hsttypes}). This is
because for redshifts $z>0.5$, the observer B-band samples the rest-frame U
and the far UV, and these are very sensitive to recent star formation. The
observer I-band, on the other hand, gives a better measure of the mass in
both the bulge and disk components.  Mock CCD images of generic Hubble
types, {\it k}-corrected to model their appearance at high redshift, demonstrate
that ordinary galaxies can look very different in the rest-frame U and UV,
becoming dominated by knots of HII emission in star forming regions which
sometimes appear as chain-like structures (Cowie \and 1996,  
Abraham {\it et al} 1996b).  

In Figure \ref{fig:dndz} we plot our predictions for the redshift
distribution of various subsamples drawn from our mock HDF catalogue. The
top panel compares the model with the Canada-France Redshift Survey (Lilly
\and 1995). Overall, the agreement is very good, although the model
produces approximately $20\%$ too many galaxies at low redshift.  This
discrepancy reflects the steep slope of the local luminosity function
($\alpha \sim -1.5$) predicted in this (and similar) models, that has led
to the claim that faint galaxies may be missing from local surveys, perhaps
due to surface brightness effects (McGaugh 1994, Frenk \and 1996).  A
detailed comparison of our model predictions with the CFRS measurement of
evolution in the galaxy luminosity function out to $z \sim 1$ is given in
Baugh \and (1996).  The middle panel shows our predicted redshift
distributions for galaxies of different morphological types with $20 <
I(F814W) < 22$, a choice that matches the limits of the sample of (Driver
\and 1995b) already selected for spectroscopic follow-up. We predict that
spirals and SOs should have similar redshift distributions, with a median
value of $z \simeq 0.45$ and a mean of $z \simeq 0.50$, whereas ellipticals should
have a shallower distribution, with a median of $z \simeq 0.25$. The redshift
distribution of irregular/merger galaxies peaks at an even lower redshift
than this but it is very flat and has an extended tail beyond $z>1$. The
exact shape of this tail depends on the choice of IMF (Cole \and 1994).
Finally, the bottom panel gives our predictions for an even fainter
magnitude slice which could, in principle, be measured with the Keck
telescope.  For galaxies with $22 < I < 24$, we predict a median redshift
of $z_{m} \simeq 0.67$, increasing to $z_{m} \simeq 0.93$ for $24 < I <
26$. This increase in median redshift with decreasing flux contrasts with
the behaviour of the `maximal merger model' of Carlberg (1996), in which
the median redshift remains around $z \simeq  0.6$ for galaxies fainter than
$I = 20$.

The remarkable agreement between our theoretical predictions and the count
data displayed in Figures~1 and~2 is not exclusive to the standard, flat
CDM cosmogony. Equally good fits to the counts are obtained, for example,
in an open CDM model with $\Omega=0.3$ (Baugh, Cole \& Frenk, in
preparation).  At high $z$, however, the redshift distributions do depend
on the cosmological parameters. For this low-$\Omega$ model they have a
similar shape to that of the standard model, but the peaks move by $\simeq 
0.3$ to higher redshift. Thus, for the open $\Omega=0.3$ cosmology, we
predict median values of $z_{m} \simeq 0.91$ and $z_{m} \simeq 0.128$ for
samples selected with $22 < I < 24$ and $24 < I < 26$ respectively. Our
predicted redshift distributions are in rough agreement with current
observational data. Preliminary results by Koo \and (1996) give a
median redshift of $z \simeq 0.81$ for a sample fainter than $ I = 22$. Cowie
\and (1995) find that 40 out of 281 objects identified as galaxies in a
sample with $I < 22.5$ have measured redshifts $z>1$. At this flux limit,
our $\Omega=1$ model has $7\% $ of all galaxies lying at $z > 1$ and our
open $\Omega=0.3$ model has 2 to 3 times as many. 


\section{Conclusions}
  
The simplest hierarchical clustering model -- $\Omega=1$ standard CDM --
provides an acceptable theoretical framework for understanding the
variation with flux of the relative numbers of galaxies of different
morphological types seen in recent HST data. These data appear consistent
with this extreme model in which galaxy formation occurs at relatively low
redshifts (Frenk \and 1985, Baugh \and 1996). Number counts, however,
provide only a limited picture of galaxy evolution and, within our class of
models, they are compatible with a range of cosmological parameters.
Essential complementary information is furnished by spectroscopic surveys
to very faint limits such as those of Lilly \and (1995) and Ellis \and
(1996). In an earlier paper (Baugh \and 1996) we showed that, in broad
terms, these data can
also be successfully interpreted in the context of these
models. Preliminary calculations (Baugh, Cole, Lacey \& Frenk 1996 in
preparation) suggest, further, that the existence of a population of
``Lyman-break'' galaxies at $z\simeq 3$, recently established by Steidel \and
(1996), may be compatible even with our standard $\Omega=1$ cosmology.  In this
paper we have presented model predictions for spectroscopic follow-up
programmes of the HST photometric surveys which offer the prospect of
discriminating between different cosmologies.  More generally, these and
related high-redshift studies, will soon provide a definitive test of the
idea that galaxies formed by the hierarchical aggregation of
gravitationally unstable primordial fluctuations.

\section*{Acknowledgements}
We would like to thank Bob Abraham, Simon Driver, Karl Glazebrook and Nigel
Metcalfe for providing counts data in electronic form and Stephane Charlot
for making available a revised version of the Bruzual and Charlot stellar
population models.  CMB acknowledges a PPARC research assistantship and SMC
a PPARC advanced Fellowship.  This work was supported in part by a PPARC
Rolling Grant.

\vspace{1cm}

\setlength{\parindent}{0cm}

{\bf References} 

\small

\newcommand{\mn}{{\em Mon. Not. R. astr. Soc}}
\newcommand{\apj}{{\em Astrophys. J.}}
\newcommand{\apjs}{{\em Astrophys. J. Suppl.}}
\newcommand{\aj}{{\em Astron. J.}}
\renewcommand{\aa}{{\em Astr. Astrophys.}}
\newcommand{\ass}{{\em Astrophys. Space Sci.}}
\newcommand{\nat}{{\em Nature}}
\def\refe {\par \hangindent=.7cm \hangafter=1 \noindent}

\refe
Abraham, R.G., Tanvir, N., Santiago, B.X., Ellis, R.S., 
Glazebrook, K., van den Bergh, S., 1996a, \mn\, 279, L47 

\refe
Abraham, R.G., van den Bergh, S., Glazebrook, K., Ellis, R.S., 
Santiago, B.X., Surma, ?., Griffiths, R.G., 1996b, \apjs\, in press

\refe
Barnes, J.E., Hernquist, L., 1992, 
Ann. Rev. Astron. Astroph., 30, 705

\refe
Baugh, C.M., Cole, S., Frenk, C.S., 1996 \mn\, submitted

\refe
Bond, J.R., Cole, S., Efstathiou, G., Kaiser, N., 
1991, \apj\, 379, 440

\refe
Bower, R.G., Lucey, J.R., Ellis, R.S., 1992, \mn\, 254, 589

\refe
Bruzual, G., Charlot, S., 1993, \apj\, 405, 538

\refe
Butcher, H.R., Oemler, A., 1984, \apj 285, 426

\refe
Carlberg, R.G., 1996, in {\it Galaxies in the Young Universe}, 
ed H. Hippelein, (Springer) in press

\refe
Casertano, S., Ratnatunga, K.U., Griffiths, R.E., Im, M., 
Neuschaffer, L.W., Ostrander, E.J., Windhorst, R.A., 1995, 
\apj\, 453, 599

\refe
Charlot, S., Worthey, G., Bressan, A., 1996, \apj\, 457, 625

\refe
Clements, D.L., Sutherland, W.J., McMahon, R.G., Saunders, W., 1996, 
\mn\, 279, 477

\refe
Cole, S., Kaiser, N., 1988, \mn\, 233, 637

\refe
Cole, S., 1991, \apj\, 367, 45

\refe
Cole, S., Arag\'{o}n-Salamanca, A., Frenk, C.S., 
Navarro, J.F., Zepf, S.E., 1994, \mn\, 271, 781

\refe
Cowie, L.L., Hu, E., Songailia, A., 1995, Nature, 377, 603-606

\refe
Cowie, L.L., Hu, E., Songailia, A., 1996, \apj\, in press

\refe
Dressler, A., 1980 \apj 236 351

\refe
Driver, S.P., Windhorst, R.A., Ostrander, E.J., Keel, W.C., 
Griffiths, R.E., Ratnatunga, K.U., 1995a, \apj\,  449, L23

\refe
Driver, S.P., Windhorst, R.A., Griffiths, R.E., 1995b, \apj\, 453, 48

\refe
Ellis, R.S., Colless, M., Broadhurst, T., Heyl, J., 
Glazebrook, K., 1996, \mn\, 280, 235

\refe
Frenk, C.S., White, S.D.M., Efstathiou, G., and Davis, M., 1985, 
{\em Nature}, 317, 595.

\refe
Frenk, C.S., Baugh, C.M., Cole, S., 1996 
IAU symp. 171, pp 247, {\it New light on Galaxy 
Evolution}, eds. Bender, R.,  Davies, R.L., (Kluwer)

\refe
Glazebrook, K., Ellis, R., Santiago, B., Griffiths, R., 1995, 
\mn\, 275, L19   

\refe 
Heyl, J.S., Cole, S., Frenk, C.S., Navarro, J.F., 1995, \mn\, 274, 755

\refe
Kauffmann, G., White, S.D.M., Guiderdoni, B., 1993,
 \mn\, 264, 201

\refe
Koo, D.C \and 1996 \apj\, in press

\refe
Lacey, C., Silk, J., 1991, \apj\, 381, 14

\refe 
Lacey, C., Guiderdoni, B., Rocca-Volmerange, B., Silk, J., 1993, \apj\, 
402, 15

\refe
Lilly, S.J., Tresse, L., Hammer, F., Crampton, D., 
Lefevre, O., 1995, \apj\, 455, 108

\refe
McGaugh, S., 1994, {\em Nature}, 367, 538

\refe
Metcalfe, N., Shanks, T., Campos, A., Fong, R., Gardner, 
J.P., 1996, {\em Nature} submitted

\refe
Mihos, J.C., Hernquist, L., 1994a, \apj\,  425, L13

\refe
Mihos, J.C., Hernquist, L., 1994b, \apj\,  431, L9

\refe
Miller, G.E., Scalo, J.M., 1979, \apjs\, 41, 513

\refe
Navarro, J.F., White, S.D.M., 1993, \mn\, 265, 271

\refe
Navarro, J.F., Frenk, C.S., White, S.D.M., 1995, 
\mn\, 275, 56

\refe
Scalo, J.M., 1986, {\it Fundam. Cosmic Physics,} 11, 1

\refe
Steidel, C., Giavalisco, M., Pettini, M., Dickinson, M., Adelberger, K.L., 
1996 \apj 462, L17

\refe
White, S.D.M., Frenk, C.S., 1991, \apj\, 379, 52

\end{document}